\documentclass[conference]{IEEEtran}
\usepackage{fancyhdr}
\usepackage{graphicx, booktabs}
\usepackage{algorithm}
\usepackage{algpseudocode}
\usepackage{hyperref}
\usepackage{subcaption}
\usepackage{placeins}
\usepackage{caption}
\usepackage{amsmath}
\usepackage{tikz}
\usepackage{listings}
\algrenewcommand\algorithmicrequire{\textbf{Input:}}
\algrenewcommand\algorithmicensure{\textbf{Output:}}

\title{LLBPE: Linked-List Based \\ GPU-Parallel BPE Tokenizer}

\author{
    \IEEEauthorblockN{Aditya Kovilur}
    \IEEEauthorblockA{aditya373@icloud.com \\ Texas A\&M University \\ College Station, TX, USA}
    \and
    \IEEEauthorblockN{Varun Chandra Shekar}
    \IEEEauthorblockA{vcs@tamu.edu \\ Texas A\&M University \\ College Station, TX, USA}
    \and
    \IEEEauthorblockN{Ariful Azad}
    \IEEEauthorblockA{ariful@tamu.edu \\ Texas A\&M University \\ College Station, TX, USA}
}

\begin{document}

\maketitle

\begin{abstract}
Every LLM inference begins with tokenization, which converts raw input bytes into the discrete token sequence the model consumes. For text, this step is often implemented using Byte Pair Encoding (BPE), an algorithm originally introduced for data compression. BPE has traditionally run on the CPU with extensive optimization, but recent work has moved it to the GPU for higher throughput. We show that these GPU implementations are bottlenecked not by computation but by data movement. We develop LLBPE that represents the token sequence as an array-based linked list so that each merge reduces to a constant-time pointer update. Furthermore, LLBPE fuses rank lookup, minimum selection, and merging into a single kernel to eliminate redundant hash map queries. LLBPE achieves up to 5.2× higher throughput than the best existing GPU implementation and 24.6× over optimized CPU implementations, at the cost of minor discrepancies in tokenized output.
\end{abstract}


\section{Introduction}
Large Language Models (LLMs) sit at the center of contemporary machine learning, and their end-to-end inference pipelines have received extensive optimization. 
Every such pipeline begins with the same step. 
Raw input is split into discrete units called \emph{tokens} before the model can process it. 
Although tokenization is not the dominant runtime cost of an individual forward pass, it is executed on every request, and its overhead accumulates significantly in high-volume serving environments.
Accelerating tokenization is therefore critical for maximizing overall pipeline throughput.

Given its central role in natural language processing, various tokenization schemes have been developed, including WordPiece~\cite{wu2016googlesneuralmachinetranslation}, SentencePiece~\cite{kudo-richardson-2018-sentencepiece}, and Unigram~\cite{Raffel2019}.
Among these, Byte-Pair Encoding (BPE)~\cite{gage1994new,sennrich2016neural} has emerged as the standard for modern LLMs, powering the GPT family~\cite{NEURIPS2020_1457c0d6} and leading open-weight models such as Llama~\cite{grattafiori2024llama3herdmodels}.
Consequently, accelerating BPE through GPU parallelism is of high practical importance.

BPE tokenization has traditionally relied on highly optimized CPU implementations, such as OpenAI's tiktoken~\cite{tiktoken} and Hugging Face Tokenizers~\cite{huggingface}.
However, CPU-based approaches are fundamentally constrained by host core counts and memory transfers.
Recent efforts have sought to offload BPE to the GPU.
BlockBPE~\cite{you2025blockbpeparallelbpetokenization} parallelizes merge passes within CUDA thread blocks, while GPUTOK~\cite{kadamba2026gputokgpuacceleratedbyte} maps exact BPE merge rules directly to GPU hardware.
Nevertheless, both methods maintain a flat array as their underlying sequence representation, severely limiting their performance scalability.

The primary performance bottleneck in current GPU algorithms lies in excessive data movement.
In a flat array, merging two adjacent tokens requires closing the resulting gap, either by shifting all downstream elements (as in BlockBPE) or by copying the sequence to an auxiliary buffer (as in GPUTOK).
Consequently, each merge incurs an $O(N)$ memory overhead relative to sequence length ($N$).
On a GPU, these global memory writes are far more expensive than the hash map lookups and
rank comparisons that constitute the actual work of BPE.

We address this with Linked-List BPE (LLBPE), a parallel BPE algorithm that represents the token sequence as an array-based linked list rather than a flat array.
In this setting, a merge becomes a constant-time pointer update touching only the two tokens being merged, leaving every other position untouched. 
We further fuse rank lookup,
minimum selection, and merging into a single kernel so that each hash
map value is retrieved once and reused.
On our set of parameters, LLBPE achieves a peak throughput of 370~MB/s, a 5.2$\times$ improvement over GPUTOK at
71.3~MB/s, while introducing only minor discrepancies along thread-block boundaries.

This trade-off suits a growing class of deployments. 
As LLM serving diversifies into low-latency and high-volume configurations alongside deep reasoning workloads, throughput-sensitive settings increasingly prioritize faster processing at every pipeline stage over exact output fidelity. 
The same applies under heavy request loads, where sustaining throughput matters more than maximizing model accuracy.

We summarize our main contributions as follows:
\begin{itemize}
    \item \textbf{Bottleneck Analysis:} We identify global data movement in flat-array token representations as the primary efficiency bottleneck in existing GPU BPE implementations.
    \item \textbf{LLBPE Algorithm:} We propose an array-based linked-list representation that reduces per-merge data movement from $O(N)$ to $O(1)$, and fuse candidate selection and merging into a single unified GPU kernel.
    \item \textbf{Empirical Evaluation:} We evaluate LLBPE against CPU baselines (tiktoken, Hugging Face Tokenizers) and GPU implementations (BlockBPE, GPUTOK). LLBPE achieves peak speedups of $5.2\times$ over prior GPU methods and $24.6\times$ over CPU-based implementations.
    \item \textbf{Implementation}: LLBPE is publicly available \href{https://github.com/SPARTACUS5329/LLBPE}{here}. The setup is mentioned in the Artifact Description section in the appendix.
\end{itemize}

\section{Background}
\subsection{Byte Pair Encoding}
Byte Pair Encoding (BPE) was introduced by Gage~\cite{gage1994new} as a data compression algorithm. 
Given an input byte sequence, BPE repeatedly finds the most frequent adjacent pair of bytes, replaces every occurrence of that pair with a single unused byte, and records the substitution in a lookup table.
This process repeats until no pair occurs more than once. The original sequence can be recovered by replaying the substitutions in reverse order.

As an example, consider the input string \texttt{aaabdaaabac}.
The pair \texttt{aa} is the most frequent, so it is replaced by an unused symbol \texttt{Z}, yielding \texttt{ZabdZabac} with the table entry $\texttt{Z} {\rightarrow} \texttt{aa}$. 
Next, the pair \texttt{ab} is most frequent. Replacing \texttt{ab} with \texttt{Y} produces \texttt{ZYdZYac}
with the additional entry $\texttt{Y} \rightarrow \texttt{ab}$. 
The process continues until no pair appears more than once.

Sennrich et al.~\cite{sennrich2016neural} adapted BPE for subword tokenization in neural machine translation. 
It has since become the dominant tokenization method in large language models including GPT-4~\cite{openai2023gpt4} and LLaMA~\cite{touvron2023llama}. 

\subsection{Training and Encoding Phases}
The use of BPE in LLMs typically involves two phases.
In an offline \emph{training phase}, a fixed-size vocabulary of merge rules is learned from a training corpus.
Starting from a character-level (or byte-level) vocabulary, the algorithm repeatedly counts all adjacent token pairs across the corpus, merges the most frequent pair into a new token, and appends the rule to an ordered
merge table. 
This repeats for a predetermined number of merges (e.g., 50{,}257 tokens for GPT-2 or 100{,}256 for GPT-4's \texttt{cl100k\_base} vocabulary).
The goal of this training phase is to learn an ordered merge table that can be used to tokenize arbitrary text at inference time. 

In the \emph{encoding (inference) phase}, the tokenizer does not re-count frequencies of new input strings. 
Instead, it splits the input into bytes and iteratively scans for adjacent pairs that appear in the learned merge table.
At each step, the pair with the lowest index (highest priority) in the ordered merge table is merged, regardless of how many times it occurs in the current input.
This process repeats until no adjacent pair in the current sequence matches any entry in the merge table.
Because every single byte is always a valid token in the vocabulary, BPE guarantees that any input can be encoded.
This paper focuses exclusively on parallelizing the BPE encoding phase on the GPU, taking a pre-trained merge table as a fixed input. 

\subsection{Pre tokenization}
Modern tokenizers such as OpenAI's tiktoken and HuggingFace Tokenizers also apply a regex-based pre-tokenization step before BPE merging. This step splits the input at word boundaries, whitespace, and punctuation using patterns such as GPT-2's splitting regex, ensuring that merges do not cross word boundaries. 
Pre-tokenization typically accounts for a substantial fraction of total tokenization time and is one of the key challenges in moving BPE to the GPU, since regex evaluation involves backtracking and branching logic that resists parallelization.

% BPE (Byte Pair Encoding) was introduced as a compression algorithm to make large strings more compact by merging sub-strings together repeatedly to generate a much smaller output that represents the input using a pre-computed map from string IDs to their values~\cite{gage1994new}. 
% This algorithm has been adopted in the LLM space to tokenize the input data for models to preserve contextual value from the input and preventing them from generating non-existent context by learning the relation between two parts of an input that would otherwise not be related. BPE works in the following steps given the input string:

% \begin{enumerate}
%     \item Find the byte pair that occurs most frequently
%     \item Replace the identified pair with a new unused byte
%     \item Encode a table with this mapping from the new byte to the old pair
%     \item Repeat this process until each pair has the same frequency
% \end{enumerate}

% At the end of this algorithm, the input string would be compressed to a new sequence of bytes that can be easily decompressed using the mapping table. Taking inspiration from this, LLMs use tokenization to merge tokens repeatedly until they do not find a possible merge. These merges are looked up from a table which is constructed during the training phase. The only tweak that the LLM tokenizer makes is in \textit{Step 1}, where it chooses the pair with the highest merge priority instead of the pair that occurs the most. The merge priority is stored as part of the pre-computed table generated during the training phase.
\section{Related Work}
\subsection{Tokenization Methods}
Subword tokenization is a prerequisite for all modern LLMs, and three
methods dominate in practice.
BPE is the most widely used tokenization method in modern LLMs. 
It is used throughout the GPT family~\cite{NEURIPS2020_1457c0d6}.
Recent open-weight models such as Llama~3~\cite{grattafiori2024llama3herdmodels} also replaced the SentencePiece-BPE scheme of Llama~1/2 with a larger, tiktoken-style byte-level BPE vocabulary.

WordPiece is used by BERT-family models~\cite{devlin2019bert}.
Unlike BPE, WordPiece performs a left-to-right greedy longest-match search against a learned vocabulary, consuming the longest recognized token at each position before advancing. 
It does not use a global merge priority and does not consider pairs at arbitrary positions in the string.

SentencePiece is a tokenization
framework that supports both BPE and Unigram Language Model~\cite{kudo2018subwordregularizationimprovingneural} backends. 
They use a single Viterbi (dynamic programming) pass to find the segmentation that maximizes the probability of the token distribution.
The entire algorithm runs as one dynamic programming pass over the corpus, without a merging step at each iteration.
Models such as T5~\cite{Raffel2019} and
ALBERT~\cite{lan2020albertlitebertselfsupervised} use SentencePiece with the Unigram backend.

%which is Viterbi-search-based rather than merge-based: the core logic
%is to find the segmentation of the entire corpus that maximizes the probability of the token distribution.

% \textit{WordPiece}~\cite{wu2016googlesneuralmachinetranslation} was introduced for \textit{BERT}~\cite{devlin2019bert} style LLMs and it is based on a greedy search for the longest possible token within the corpus from the beginning of the string. The tokens are learned while training on the basis of a threshold on the vocabulary size. During inference, the corpus is linearly scanned for the longest possible tokens from the start of the string, unlike BPE which merges at any position in the entire string which has the highest merge priority.

\subsection{CPU-Based Implementations}
Several optimized CPU implementations of BPE encoding exist in the literature.
HuggingFace Tokenizers offers a Rust-based library supporting BPE, WordPiece, and Unigram backends with Python bindings. 
OpenAI's tiktoken optimizes BPE encoding in Rust. 
fastBPE~\cite{fastbpe} provides a multithreaded C++ implementation with memory-mapped I/O.

\subsection{GPU-Accelerated Tokenization}
% Because its merge order is governed by a single, static rank table, byte-level BPE exposes a per-pair, per-pass parallel structure that neither WordPiece's longest-match search nor SentencePiece's Unigram Viterbi decoding provide, making it the natural target for the GPU kernel design we present in this work.
Several recent efforts have targeted GPU acceleration of tokenization.
The RAPIDS cuDF library~\cite{rapidsai_cudf} provides a GPU-based
subword tokenizer for WordPiece in BERT pipelines.
BlockBPE~\cite{you2025blockbpeparallelbpetokenization} proposes a parallel BPE implementation that replaces regex-based pre-tokenization with byte-level splitting to enable intra-block parallelism on the GPU
GPUTOK~\cite{kadamba2026gputokgpuacceleratedbyte} builds on similar ideas to implement GPT-2's exact merge rules on the GPU.

%We adopt BPE as the target of our GPU parallelization due to its widespread use in the current generation of LLMs. 

\begin{figure*}[t]
    \centering
    \includegraphics[width=0.7\textwidth]{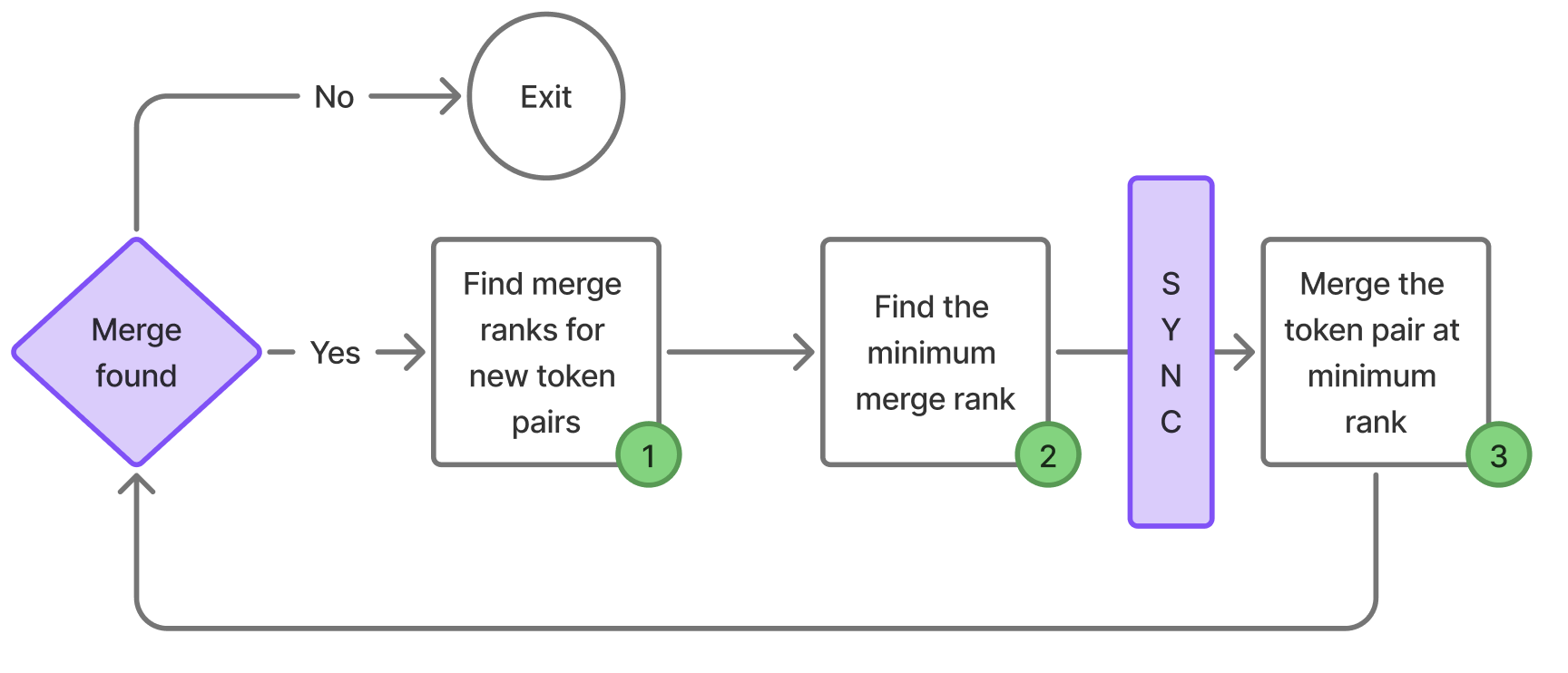}
    \caption{An overview of LLBPE within a thread block. These operations are performed independently in eack block.
    % https://www.figma.com/board/pbMJepwCwqDJH52XO4lUGN/MultiBlockBPE?node-id=0-1&t=JIFXABqjgJsLLW90-1
    }
    \label{fig:flow}
\end{figure*}
\section{Method}
\subsection{Pre Tokenization}
Before applying BPE merges, the input text is split into pre-tokens such as words and numbers.
Pre-tokenization enforces the semantic rules of the input language, thereby preserving contextual information.
For example, a number such as \texttt{1000} should always be tokenized as a single unit rather than split into \texttt{10} and \texttt{00}. 
Analogous rules govern punctuation, contractions, and similar constructs. 
Without this step, the string \texttt{sample word} could
produce tokens such as \texttt{samp}, \texttt{lewo}, \texttt{rd},
merging the unrelated characters \texttt{le} and \texttt{wo} across a
word boundary and introducing spurious context. 
Modern tokenizers implement pre-tokenization by lexing the input stream with a set of regular expressions that greedily form the longest valid pre-tokens. 

BlockBPE removes
pre-tokenization entirely to maximize GPU performance, at the cost of small deviations in tokenization output. 
Our implementation retains pre-tokenization but overlaps it with file I/O by processing the input as a stream, so that pre-tokenized spans are ready for GPU processing as soon as they are read from disk.

We implement pre-tokenization as a single linear scan over the input using a pre-computed trie that encodes the tokenizer's regular expressions using the Knuth--Morris--Pratt (KMP) algorithm~\cite{kmp-paper}.
In this approach, at each node, a valid input advances to a child node, while a mismatch follows a pre-computed failure link that falls back to the longest previously matched suffix. 
This ensures that every byte is visited at most a constant number of times. 
The time complexity of the scan is $O(N + P)$, where $N$ is the length of the input and $P$ is the maximum depth of the trie. 
In practice $N \gg P$, so the cost reduces to $O(N)$, which is the same asymptotic cost as simply copying the input into an array. 
The pre-tokenization overhead is therefore amortized over the cost of populating the token buffer that is subsequently transferred to the GPU.

% Essentially, the input is streamed linearly over the trie preserving the computation time complexity before populating the tokens array. The time complexity of the substring matching algorithm is \textit{O(N + P)} where \textit{N} is the length of the input string and \textit{P} is the maximum depth of the trie. In our case, $N >> P$, therefore the complexity can be reduced to \textit{O(N)} which would match the streamed input being stored in an array directly. The cost for the scan is amortized over populating the tokens array.

\subsection{Batching and Thread Mapping}
We implement BPE on the GPU by partitioning the input text into batches. Each batch contains \texttt{BATCH\_SIZE} (number of) substrings, and each substring is of length $N = \texttt{SEQ\_LEN}$ initial tokens (each token is a byte after pre-tokenization and before the first iteration of BPE). 
Hence, $\texttt{BATCH\_SIZE} \times \texttt{SEQ\_LEN}$ is equal to the total number of bytes processed in one GPU kernel launch.

We assign each batch to its own thread block of $N$ threads. 
As a result, each thread is responsible for exactly one token at any given point during execution.
Since block scheduling is handled by the CUDA scheduler, our algorithm centers around how each batch is processed within a thread block.
% The post pre-tokenization mapping inside the GPU is shown in \textit{Figure}~\ref{fig:mapping}, where the \texttt{SEQ\_LEN} is taken as 3 and the \texttt{BATCH\_SIZE} is taken as 2. This would mean that there 2 batches of substrings, each of length \texttt{SEQ\_LEN = 3}. For example, if the input text is \"abcdef\", the string would be mapped to the GPU threads as shown in \textit{Figure}~\ref{fig:mapping}.

Since each thread handles exactly one token and we employ no thread coarsening (in contrast to BlockBPE), sequence length and batch size are not independent:
\begin{equation}
    \texttt{SEQ\_LEN} \times \texttt{BATCH\_SIZE} = \texttt{MAX\_THREADS},
    \label{eq:max_threads}
\end{equation}
where \texttt{MAX\_THREADS} is the hardware limit on the number of threads that the GPU can spawn concurrently. Due to the limitation of tokenizing an input of \texttt{MAX\_THREADS} bytes at a time, for an input larger than that, we launch the GPU kernel multiple times. For example, if \texttt{MAX\_THREADS = 200,000} and the input text is of \texttt{SIZE = 2,100,000 Bytes}, we launch the GPU kernel \texttt{ceil(SIZE / MAX\_THREADS) = 11} times.

% \begin{figure}[t]
%     \centering
%     \includegraphics[width=0.5\textwidth]{mapping.png}
%     \caption{Example mapping of tokens in to GPU threads with sequence length = 3 and batch size = 2
%     % https://www.figma.com/board/pbMJepwCwqDJH52XO4lUGN/MultiBlockBPE?node-id=0-1&t=JIFXABqjgJsLLW90-1
%     }
%     \label{fig:mapping}
% \end{figure}

\begin{algorithm*}[!t]
\caption{BPE Merge Iteration}
\label{alg:bpe-merge}
\begin{algorithmic}[1]
\Require thread index $i$, arrays \texttt{tokens[]},
         \texttt{nextToken[]}, merge hash map $H$,
         shared variable \textit{minRP}

\Statex \Comment{\textbf{Phase 1: Rank lookup}}
\If{$\texttt{nextToken}[i] = -1$}
    \State \Return \Comment{inactive thread}
\EndIf
\State $j \gets \texttt{nextToken}[i]$
\State $\textit{value} \gets H.\text{lookup}(
       \texttt{tokens}[i],\; \texttt{tokens}[j])$
\State $\textit{candidate} \gets
       (\textit{value} \;\&\; \texttt{0xFFFFFFFF00000000})
       \;\lor\; i$
       \Comment{rank (high), position (low)}

\Statex
\Statex \Comment{\textbf{Phase 2: Atomic minimum selection}}
\State $\textit{old} \gets \textit{minRP}$
\While{$\textit{candidate} < \textit{old}$}
    \State $\textit{prev} \gets
           \textproc{atomicCAS}(\textit{minRP},\;
           \textit{old},\; \textit{candidate})$
    \If{$\textit{prev} = \textit{old}$}
        \State \textbf{break}
    \EndIf
    \State $\textit{old} \gets \textit{prev}$
\EndWhile

\Statex
\Statex \Comment{\textbf{Phase 3: Merge}}
\State \textproc{syncThreads}()
\State $\textit{mergePos} \gets \textit{minRP} \;\&\;
       \texttt{0xFFFFFFFF}$ \Comment{Lower 32 bits of \textit{minRP} stores the position}
\If{$i = \textit{mergePos}$}
    \State $\texttt{tokens}[i] \gets \textit{value} \;\&\;
           \texttt{0xFFFFFFFF}$
           \Comment{Lower 32 bits of value stores merged token ID}
    \State $\texttt{nextToken}[i] \gets \texttt{nextToken}[j]$
           \Comment{skip consumed node with $j=\texttt{nextToken}[i]$ from Phase~1}
    \State $\texttt{nextToken}[j] \gets -1$
           \Comment{mark as deleted}
\EndIf
\end{algorithmic}
\end{algorithm*}

\subsection{Overview of the Algorithm}
The core of BPE encoding is a loop that repeatedly identifies the highest-priority adjacent token pair and merges it into a single token.
In a flat array representation, each merge requires shifting all subsequent elements to close the gap.
Such data shifting is expensive on the GPU, where global memory writes are costly.
We instead emulate a linked list on the GPU, where each token stores indices to its successor.

Hence, a merge reduces to updating a constant number of index entries in this linked list.
Each merge iteration consists of three steps, illustrated in
Figure~\ref{fig:flow}.

\textbf{Step~1: Rank evaluation.} For each adjacent pair in the
current sequence, look up its merge rank in the pre-trained table.
This step takes $O(N)$ time, where $N$ is the number of active tokens.

\textbf{Step~2: Minimum selection.} Find the pair with the lowest
rank (highest priority). This can be performed via linear scan or
tree-based reduction, both effectively $O(N)$ in the number of active tokens.

\textbf{Step~3: Merge.} Replace the selected pair with a single
merged token and update the sequence. 

The key difference between our approach and prior GPU implementations lies in Step~3. BlockBPE stores tokens in a flat array and shifts all elements after the merge position left by one, making Step~3 $O(N)$.
GPUTOK uses a double-buffered array, copying all $N$ elements into a second buffer in every iteration, which is also $O(N)$. 
In LLBPE's linked-list representation, Step~3 is $O(1)$ because exactly one thread updates the merged token's value and redirects its successor pointer to skip the consumed neighbor.

\subsection{Data Structures}
LLBPE relies on two GPU-resident data structures: a hash map encoding the pre-trained merge table and a pair of parallel arrays representing the token sequence as a linked list.

{\bf Merge Tables.}
We store the merge table as a GPU-resident hash map using cuCollections, passed as a parameter to the merge kernel. 
Each entry of the hash map encodes a token pair and its merge rule in two 64-bit integers:
\begin{enumerate}
    \item Key (64-bit unsigned integer): The upper 32 bits store $\text{token}_i$ and the lower 32 bits store $\text{token}_{i+1}$, packing an adjacent pair into a single 64-bit lookup key.
    \item Value (64-bit unsigned integer): The upper 32 bits store the merge rank and the lower 32 bits store the resulting merged token ID. 
\end{enumerate}

\noindent This value layout, also used by
GPUTOK, has a useful property. 
Because the rank occupies the upper bits, a numeric minimum over raw 64-bit values is equivalent to selecting the lowest rank. 
The minimum simultaneously yields both the winning rank and the corresponding merged token ID in a single comparison, with no additional lookup required. 
Since all ranks in the merge table are distinct, ties cannot occur.

{\bf Token Sequence.}
The token sequence is represented as a linked list using two parallel arrays:
\begin{itemize}
    \item \texttt{tokens[i]}: the ID of the $i$-th token in the current sequence.
    \item \texttt{nextToken[i]}: the index of the successor of position $i$ in the \texttt{tokens} array. A value of $-1$ indicates a deleted node (consumed by a merge) or the end of a pre-token boundary. 
\end{itemize}
Each thread owns exactly one position in the linked list, accessing only its corresponding entries in the \texttt{tokens} and \texttt{next} arrays. This one-to-one mapping eliminates memory access conflicts between threads and ensures coalesced reads when thread indices are contiguous.
A merge operation updates at most three entries, leaving all other array positions untouched.

\subsection{The Merge Iteration}
Each merge iteration identifies the adjacent token pair with the lowest merge rank, merges it into a single token, and updates the linked list accordingly. 
Algorithm~\ref{alg:bpe-merge} shows the complete iteration, executed by every thread in parallel. 
Rank lookup, minimum selection, and the merge itself are fused into a single kernel so that the hash map value retrieved in Phase~1 can be reused in Phase~3 without a second query.

{\bf Phase~1: Rank lookup (lines 1--6).}
A thread is \emph{active} in a given iteration if $\texttt{nextToken}[i] \neq -1$, meaning a prior merge has not consumed its token and it has a valid successor to form a pair with.
Each active thread looks up its adjacent pair $(\texttt{tokens}[i],\; \texttt{tokens}[j])$ in the hash map, where $j = \texttt{nextToken}[i]$, and caches the returned 64-bit value. 
The thread then constructs a \emph{candidate} word by masking off the lower 32 bits of the value and substituting its own position~$i$, leaving the rank in place in the upper 32 bits.

In principle, only two pairs change between iterations (the pair formed by the newly merged token and its successor, and the pair formed by the predecessor and the newly merged token), so unchanged pairs could skip this lookup. 
As an optimization, all active threads perform the lookup during initialization.
On subsequent iterations, only the threads whose pairs were modified by the previous merge repeat it, while the rest reuse their cached values. This reduces the number of hash map queries per iteration from $O(N)$ to $O(1)$ after initialization.

{\bf Phase~2: Atomic minimum selection (lines 7--14).}
Every active thread attempts to insert its candidate into the shared-memory variable \textit{minRP} using an atomic compare-and-swap (CAS) loop.  
Because the rank occupies the upper 32 bits, a numeric comparison on the packed word is equivalent to comparing by rank and breaking ties by position. 
Since all merge ranks are distinct, the final value of \textit{minRP} uniquely identifies both the winning rank and the merge position.

Note that our reduction is asymptotically similar to the tree-based reduction used in GPUTOK which has $O(N)$ time complexity. 
However, we found that the tree-based approach introduces multiple synchronization barriers per iteration, which makes it expensive in practice. 
By contrast, our flat atomic approach avoids inter-thread synchronization entirely, relying instead on atomic operations that are efficiently serialized by the hardware.

\begin{figure}[t]
    \centering
   
    \begin{subfigure}[t]{0.49\columnwidth}
        \centering
        \includegraphics[width=\textwidth]{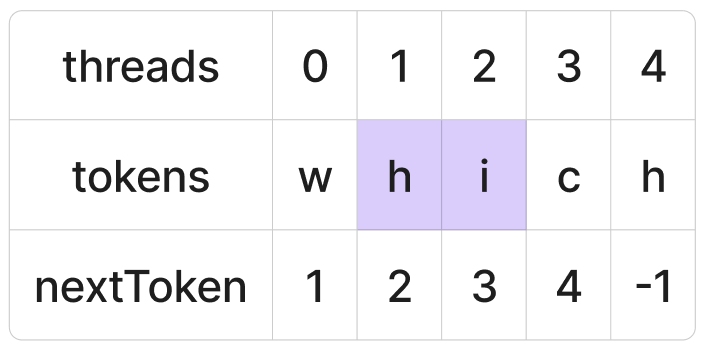}
        \caption{}
        \label{fig:merge-a}
    \end{subfigure}
    \begin{subfigure}[t]{0.49\columnwidth}
        \centering
        \includegraphics[width=\textwidth]{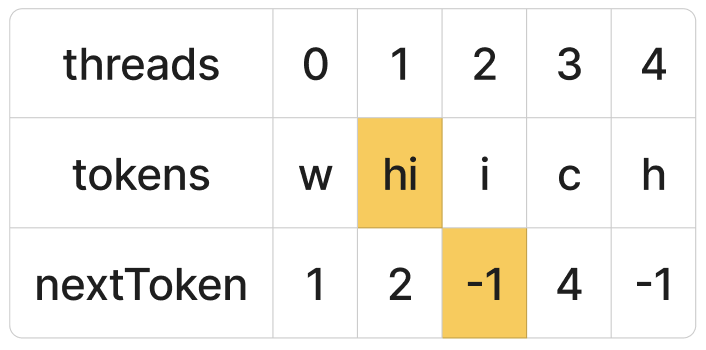}
        \caption{}
        \label{fig:merge-b}
    \end{subfigure}
     \caption{An illustration of the merge process}
    \label{fig:merge-process}
\end{figure}

{\bf Phase~3: Merge (lines 15--21).}
After a barrier synchronization, every thread reads the winning position from the lower 32 bits of \textit{minRP}. 
The thread whose position matches performs the merge.
In the merging phase, the thread extracts the merged token ID from the lower 32 bits of its cached hash map value, writes it to $\texttt{tokens}[i]$, redirects $\texttt{nextToken}[i]$ past the consumed neighbor~$j$, and sets $\texttt{nextToken}[j] \gets -1$ to mark $j$ as deleted. 
Both $j$ and the cached value are per-thread
registers held since Phase~1, so no additional memory reads or hash
map queries are needed. 
%The iteration then repeats until no adjacent pair in the sequence matches an entry in the merge table.

Figure~\ref{fig:merge-process} illustrates the merging process with an example. 
If the pair \texttt{"h"},\,\texttt{"i"} has the highest merge priority, the
two tokens merge to form \texttt{"hi"} and thread~\#2 becomes
inactive for all subsequent iterations. Only the pairs
\texttt{\{"w",\,"hi"\}} and \texttt{\{"hi",\,"c"\}} require new rank
lookups in the following iteration (Figure~\ref{fig:merge-b}).

{\bf Block-boundary handling.} 
We restrict merging within each thread block because cross-block merges require expensive inter-block synchronization through global memory barriers, which significantly reduces tokenization throughput. 
Consequently, pairs that straddle a block boundary are never merged, which can yield tokenizations that differ from a strictly serial implementation. 
Section~\ref{sec:quality} quantifies how often this occurs and its effect on tokenization quality.

\section{Experiments}
\subsection{Experimental Settings}
\paragraph{Dataset}
We evaluate the tokenizer on several corpora drawn from Tolstoy's \textit{War and Peace}\cite{wap}, Mark Twain's Tom Sawyer \cite{tom}, and Huckleberry Finn \cite{huckle}. The sizes of all these datasets are discussed in \textit{Table}~\ref{tab:dataset_sizes}.

\begin{table}[t]
    \centering
    \caption{Sizes of datasets}
    \begin{tabular}{c c}
    \toprule
         Dataset & Size  \\
         \toprule
         War and Peace & 3.2 MB \\
         Tom Sawyer &  420 KB \\
         Huckleberry Finn & 600 KB \\
         \bottomrule
    \end{tabular}
    
    \label{tab:dataset_sizes}
\end{table}

\paragraph{Platform}
All experiments were run on a GPU node of the Perlmutter supercomputer at NERSC.  
Specifically, we used an NVIDIA A100 GPU with 80GB of memory, paired with an AMD EPYC 7763 CPU (64 cores, 256GB DDR4 DRAM). Perlmutter uses a PCIe 4.0 interconnect for the CPU-GPU data transfer with a bidirectional bandwidth of 64GBps. The A100 supports approximately 200K concurrently resident threads, which upper-bounds the product of sequence length and batch size in our experiments.

\paragraph{Configuration}
We use CUDA 12.9 and gcc-native 12.3 for compilation. We also use cmake 3.30.4 for compiling cuCollections with the corresponding version of CUDA.
We use the GPT-2 \cite{gpt2vocab} tokenization vocabulary and merge table for all baselines and LLBPE.

\paragraph{Baselines}
We compare against tiktoken and HuggingFace. These are CPU-based implementations; hence, we run them on the AMD EPYC 7763 CPU as mentioned above. BlockBPE and GPUTOK are GPU implementations and are tested on NVIDIA A100. 
As there is no publicly available implementation of the original BlockBPE paper, we use the BlockBPE implementation from GPUTOK.
Tiktoken, HuggingFace, and BlockBPE are Rust implementations with Python interfaces. 
GPUTOK and LLBPE are C++ implementations.

% \paragraph{Evaluation Metrics}
% \Ariful{Discuss evaluation metrics}

% \subsection{Accuracy}

% Since our aim is not to improve tokenization accuracy, we instead evaluate how closely our tokenizer's output matches that of the trusted HuggingFace BPE tokenizer, which we adopt as the reference. The goal is to show that LLBPE tokenization aligns closely with the HuggingFace BPE tokenizer.
% Following the methodology of \cite{you2025blockbpeparallelbpetokenization}, we quantify this similarity between the two tokenizers' outputs by computing the normalized Levenshtein distance between our output and that of HuggingFace. 

\begin{figure}[t]
    \centering
    \includegraphics[width=0.5\textwidth]{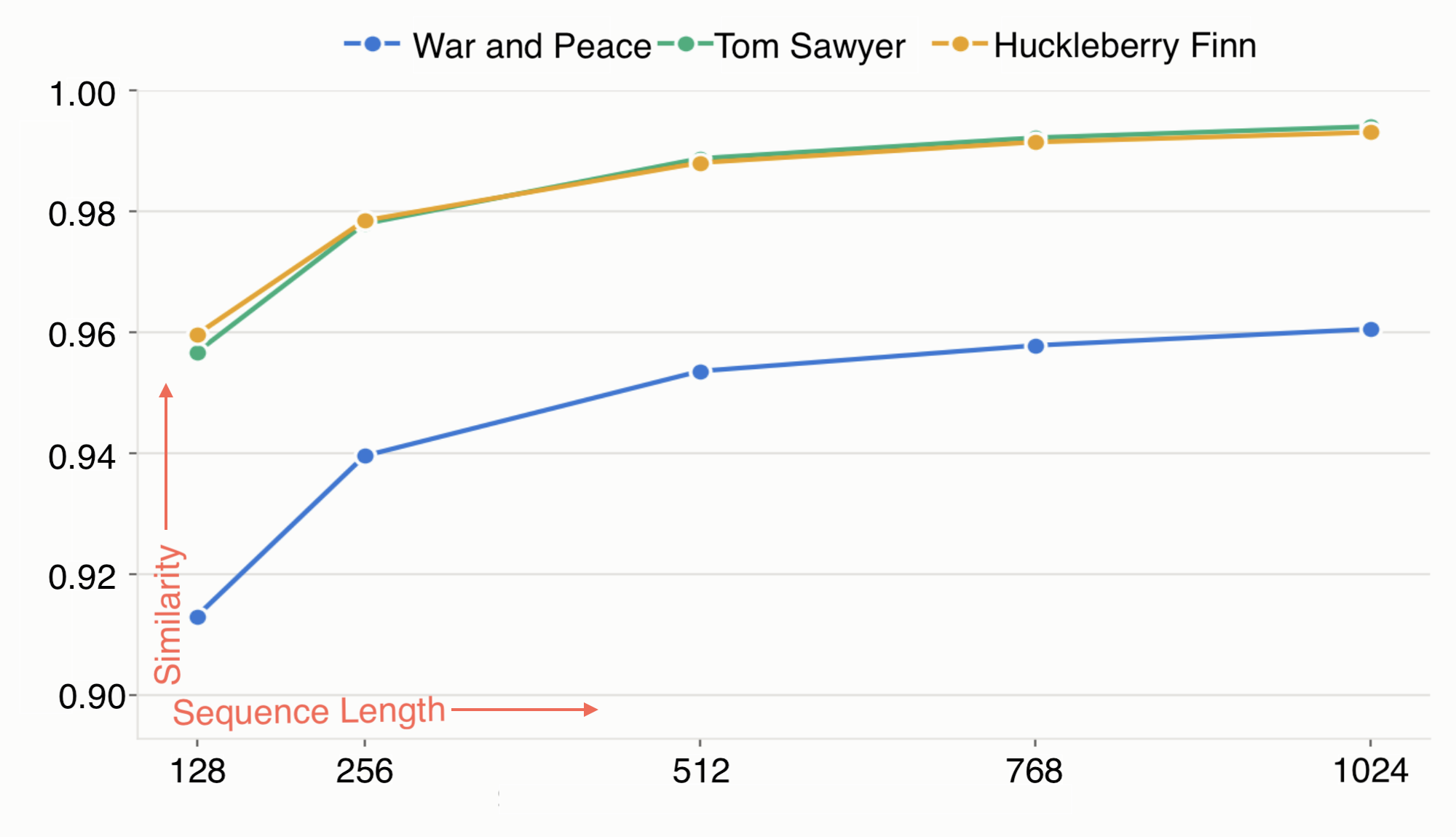}
    \caption{Comparison of similarity scores for LLBPE and HuggingFace outputs.}
    \label{fig:accuracy}
\end{figure}

\subsection{Tokenization Quality}
\label{sec:quality}
Since our aim is not to improve tokenization accuracy, we instead evaluate how closely our tokenizer's output matches that of the trusted HuggingFace BPE tokenizer, which we adopt as the reference. 
The goal is to show that LLBPE tokenization matches closely with the HuggingFace BPE tokenizer.

Following the methodology of BlockBPE, we quantify the similarity between two tokenizers' outputs by computing the normalized Levenshtein distance between our output and that of HuggingFace. For each input string $s_i$, the similarity score is defined as:
\begin{equation}
    \text{sim}(s_i) = 1 - \frac{d_L\!\left(t_i^{\text{HF}},\, t_i^{\text{LL}}\right)}{|s_i|},
    \label{eq:similarity}
\end{equation}
where $d_L(t_i^{\text{HF}}, t_i^{\text{LL}})$ is the Levenshtein distance~\cite{levenshtein1966binary} between the HuggingFace tokenization $t_i^{\text{HF}}$ and the LLBPE tokenization $t_i^{\text{LL}}$ for string $s_i$, and $|s_i|$ denotes the number of tokens in $s_i$. 
Here, the Levenshtein distance measures the minimum number of single-token insertions, deletions, and substitutions required to transform one token sequence into the other, providing a standard measure of edit-based dissimilarity between two sequences. 
A similarity score of 1 indicates that the two tokenizations are identical, while lower values indicate a greater number of edits needed to reconcile the two outputs.
The final similarity is reported as the mean of $\text{sim}(s_i)$ over all chunks in the corpus. 

Figure~\ref{fig:accuracy} shows that LLBPE is comparable to HuggingFace BPE, with similarity scores close to 99\% for \textit{Tom Sawyer} and \textit{Huckleberry Finn}, and 96\% for \textit{War and Peace}. We note that these differences do not indicate inferior accuracy on the part of LLBPE. 
They simply mean that LLBPE generates slightly different tokens than HuggingFace does.

The reason for these differences is that token pairs lying at block boundaries are not compared across blocks when evaluating merge rank in order to ensure that the algorithm operates strictly within a block for high performance. Consequently, certain merges are missed whenever a boundary token pair has the highest merge priority, introducing a discrepancy between the algorithm's output and the ground-truth tokenization. This does not cause any data loss, as it merely leaves some tokens split at the block boundary.

\begin{figure}[!t]
    \centering
    \includegraphics[height=2in]{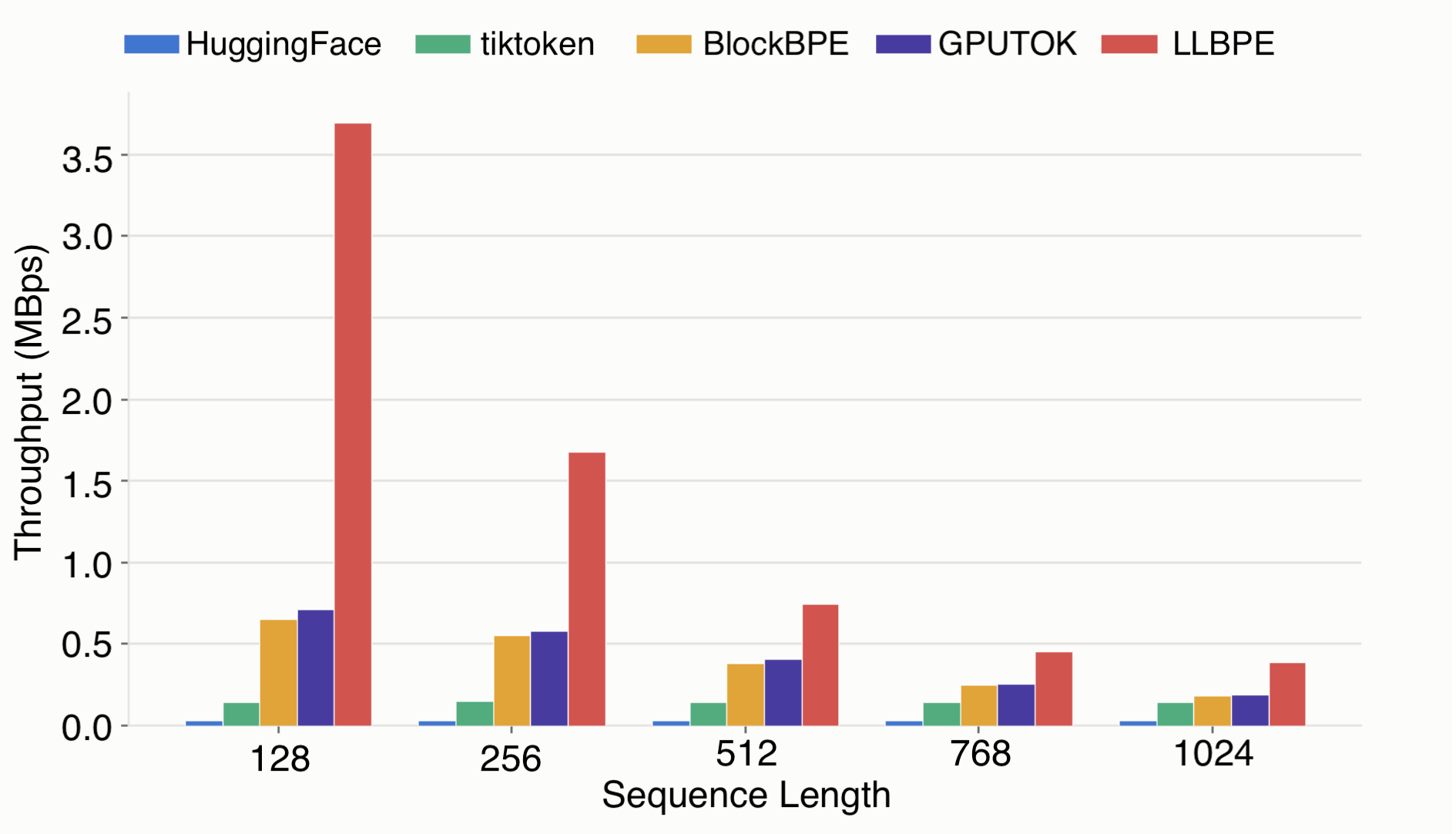}
    \label{fig:peak-seq}
    \caption{Peak throughput comparison of LLBPE with tiktoken, HuggingFace, BlockBPE and GPUTOK over varying sequence lengths.}
    \label{fig:throughput}
    \label{fig:peak-throughput}
\end{figure}

Figure~\ref{fig:accuracy} also reveals that similarity improves with a higher \textit{sequence length}.
This is expected because a higher sequence length results in a smaller \textit{batch size} for a fixed thread budget. This corresponds to fewer blocks on the GPU and consequently fewer block boundaries, reducing the opportunities for merges to be missed at those boundaries.

% The accuracy is a function of the dataset used and the observed maximum accuracy is reported in \textit{Table} \ref{tab:max-accuracy}. The minimum for \textit{War and Peace} \cite{wap} is 91.3\%. While the tokenized output does not exactly match the ground truth, there is no data loss and it is due to a token at the block boundary being split into parts.

% \begin{table}[t]
%     \centering
%     \begin{tabular}{lc}
%         \hline
%         \textbf{Corpus} & \textbf{Max Similarity} \\
%         \hline
%         War and Peace       & 0.9606 \\
%         Tom Sawyer          & 0.9941 \\
%         Huckleberry Finn    & 0.9931 \\
%         \hline
%     \end{tabular}
%     \caption{Maximum Similarity Score Achieved per Corpus}
%     \label{tab:max-accuracy}
% \end{table}

\subsection{Tokenization Throughput}
We define throughput as the number of bytes a tokenizer processes per unit time, reported in MB/s. Formally, given a corpus of total size $B$ bytes and a measured runtime $t$ (in seconds), throughput is computed as 
\begin{equation}
    \text{Throughput (MB/s)} = \frac{B}{10^6 \times t},
    \label{eq:throughput}
\end{equation}
where $B$ denotes the total number of bytes in the corpus and $t$ denotes the runtime of the program.

We compare LLBPE with two widely used CPU baselines, tiktoken and Hugging Face, and two recently developed GPU BPE implementations, BlockBPE and GPUTOK. As discussed previously, the product of batch size and sequence length equals the maximum number of threads in LLBPE. Consequently, as sequence length increases, batch size decreases proportionally. Figure~\ref{fig:throughput} presents the throughput results across various sequence lengths.

All reported numbers in \textit{Figure}~\ref{fig:peak-throughput} correspond to the merge kernel, which does not include the file I/O required for reading the corpus.
The GPU-based baselines skip pre-tokenization, hence the results shown are a direct comparison between the merge kernels. However, the CPU-based baselines run pre-tokenization as the first step on a string buffer passed as input after the corpus is read. Since we perform pre-tokenization on a streamed input, we were unable to show a direct comparison of the effect of pre-tokenization on the end to end execution time with the CPU-based baselines.

\paragraph{Overall throughput} Figure~\ref{fig:throughput} shows that the GPU tokenizers, LLBPE, BlockBPE, and GPUTOK, consistently perform much better than the highly-optimized CPU tokenizers from OpenAI and HuggingFace, across all batch sizes and sequence lengths. This is not surprising, given the massive parallelization available on the GPU compared to multicore CPUs. Between the CPU baselines, tiktoken achieves higher throughput than HuggingFace across all experiments. Between the two GPU baselines, GPUTOK slightly outperforms BlockBPE. However, across all experiments, LLBPE performs significantly better than both GPUTOK and BlockBPE. In the best case (sequence length 128), LLBPE achieves a throughput of 370 MB/s, which is $7\times$ faster than GPUTOK and $\sim20\times$ faster than tiktoken. In the worst case (sequence length 1024), LLBPE achieves 38.7 MB/s, which is $2\times$ faster than GPUTOK and $\sim2.5\times$ faster than tiktoken. Hence, LLBPE performs the best overall across all configurations explored in this paper.

\paragraph{Throughput with respect to Sequence length}
Sweeping across various sequence lengths, we notice much higher throughput values for lower sequence lengths and higher batch sizes. The difference between the throughputs at sequence lengths 128 and 1024 is an order of magnitude, and this can be attributed to the warp scheduling having a lower overhead for shorter sequence lengths. 
The most expensive step of LLBPE is the global memory access for merge ranks, which triggers a warp scheduler to hide the latency caused by it. 
A shorter sequence length corresponds to a smaller block size, \textit{i.e.} a sequence length of 128 would correspond to 128 threads in each GPU block. Consequently, it has a direct impact on the number of warps within a block. Therefore, the shorter the sequence length, the fewer the warps within a GPU, which leads to easier latency-hiding scheduling between warps. 
This is because block scheduling takes precedence once all warps are idle, which happens sooner if there are fewer warps.

\section{Conclusion}
We presented LLBPE, a GPU-accelerated BPE tokenizer that achieves a
peak throughput of 370~MB/s (with our configuration), a 5.2$\times$ improvement over the next
best baseline, GPUTOK, at 71.3~MB/s. 
The performance gain follows from a single design principle. 
Because BPE encoding involves only hash map lookups and rank comparisons, its cost on the GPU is dominated by data movement rather than computation. 
LLBPE minimizes this movement by representing the token sequence as an array-based linked list, reducing each merge from an $O(N)$ array shift to an $O(1)$ pointer update.
Additionally, LLBPE fuses rank lookup, minimum selection, and merging into a single kernel so that hash map values retrieved once are reused without re-querying global memory.

Although tokenization is not the dominant cost of LLM inference, it
is performed on every input and at high frequency in serving
pipelines. Accelerating it on the GPU also keeps token sequences
resident in device memory, avoiding host-to-device transfers in
workloads where the downstream model already runs on the GPU.

LLBPE has one notable limitation. Merges are confined to thread block
boundaries, so pairs straddling adjacent blocks are never merged.
This introduces minor discrepancies relative to the reference
HuggingFace tokenizer output, and in this work, we do not study the impact of this difference in the downstream tasks of LLMs.
Extending LLBPE to resolve cross-block merges without sacrificing throughput, for example through a lightweight boundary fix-up pass, is a natural direction for future
work.

\section*{Acknowledgments}
This research was funded in part by DOE grants DE-SC0022098 and DE-SC0023349 and by NSF grants PPoSS CCF 2316233 and OAC-2339607. Claude AI was utilized for grammar enhancement.

\bibliography{references}
\appendix
\paragraph{Abstract}
This artifact contains the implementation for LLBPE in C++. We provide instructions on generating results for \textit{Figures}~\ref{fig:accuracy}, \ref{fig:peak-throughput}.

\paragraph{Artifact Description}: The code for \textit{LLBPE} can be found \href{https://github.com/SPARTACUS5329/LLBPE}{here}. The testing was done on an NVIDIA A100 GPU on NERSC's Perlmutter on an interactive GPU node. The setup instructions for our platform are mentioned below and can also be found in \texttt{README.md} in the root directory.

\textbf{Note:} The project can be compiled using the script in \texttt{scripts/build.sh}. The executable generated by running the build script is present in \texttt{build/} and can be run with the following arguments:
\texttt{}.

\begin{lstlisting}[language=bash]
./build/LLBPE <input-file> \
    <output-file|stdout> <mode(SEQ|BATCH)> \
    <mode-value>
\end{lstlisting}

\begin{enumerate}
    \item input-file: File address of the corpus to tokenize
    \item output-file: File address to store the tokenized output
    \item mode: \texttt{SEQ}/\texttt{BATCH} to specify which parameter is fixed
    \begin{enumerate}
        \item SEQ: sequence length
        \item BATCH: batch size
    \end{enumerate}
    \item mode-value: Value of the parameter specified in <mode>
\end{enumerate}

\paragraph{Modules}
These commands and versions are specific to the platform on which we run our tests.

\begin{lstlisting}
module load cuda # CUDA 12.9
module load gcc-native/12.3
module unload cray-mpich
module load mpich
\end{lstlisting}

\paragraph{Build}
The commands provided below are included in the build script:
\begin{lstlisting}
mkdir -p build
cd ./build
cmake ..
make
\end{lstlisting}

\end{document}